\documentclass[iop]{emulateapj}
\usepackage{verbatim}
\usepackage{gensymb}
\usepackage{graphicx}
\usepackage{array}
\usepackage{booktabs}
\usepackage{threeparttable}
\usepackage{natbib}
\newcommand{\otoprule}{\midrule[\heavyrulewidth]}
\newcolumntype{+}{>{\global\let\currentrowstyle\relax}}
\newcolumntype{^}{>{\currentrowstyle}}

\begin{document}

\title{A Multi-Ringed, Modestly-Inclined Protoplanetary Disk around AA Tau}

\author{
Ryan A. Loomis\altaffilmark{1,2},
Karin I. {\"O}berg\altaffilmark{2},
Sean M. Andrews\altaffilmark{2},
Meredith A. MacGregor\altaffilmark{2}
}

\altaffiltext{1}{Corresponding author: rloomis@cfa.harvard.edu}
\altaffiltext{2}{Harvard-Smithsonian Center for Astrophysics, Cambridge, MA 02138}

\begin{abstract}
    AA Tau is the archetype for a class of stars with a peculiar periodic photometric variability thought to be related to a warped inner disk structure with a nearly edge-on viewing geometry. We present high resolution ($\sim$0\farcs2) ALMA observations of the 0.87 and 1.3~mm dust continuum emission from the disk around AA Tau. These data reveal an evenly spaced three-ringed emission structure, with distinct peaks at 0\farcs34, 0\farcs66, and 0\farcs99, all viewed at a modest inclination of 59.1$\degree\pm$0.3$\degree$ (decidedly not edge-on). In addition to this ringed substructure, we find non-axisymmetric features including a `bridge' of emission that connects opposite sides of the innermost ring. We speculate on the nature of this `bridge' in light of accompanying observations of HCO$^+$ and $^{13}$CO (J=3--2) line emission. The HCO$^+$ emission is bright interior to the innermost dust ring, with a projected velocity field that appears rotated with respect to the resolved disk geometry, indicating the presence of a warp or inward radial flow. We suggest that the continuum bridge and HCO$^+$ line kinematics could originate from gap-crossing accretion streams, which may be responsible for the long-duration dimming of optical light from AA Tau.
\end{abstract}

\section{Introduction}
    Protoplanetary disks are expected to undergo dramatic morphological changes concurrent with the processes of planet formation. Transition disks have been suggested to be observational signatures of this evolution, with their heavily depleted inner dust cavities possibly cleared by nascent gas-giant planets \citep[e.g.][]{Papaloizou_2007, Zhu_2011}. Observations have uncovered disks with both fully and partially depleted cavities, often with remnants of an inner disk near the host star \citep[e.g.][]{Calvet_2002, Calvet_2005, DAlessio_2005, Najita_2007, Espaillat_2010, Espaillat_2011, Andrews_2011}. Recent observations, however, have revealed more complex disk dust structures. Multiple dust rings have been imaged in the disks around HL Tau, TW Hya, HD 163296, and HD 169142 \citep{ALMA_2015, Andrews_2016, Isella_2016, Fedele_2017}, and visibility modeling has suggested one other candidate system, DM Tau \citep{Zhang_2016}. These rings, and the gaps between them, may trace planet formation at its earliest stages \citep[e.g.][]{Flock_2015, Ruge_2016}, although other explanations have also been proposed \citep[e.g.][]{Zhang_2015a, Okuzumi_2016}. 
    
    At small scales, the disk around the low mass (0.85 M$_{\odot}$) T Tauri star AA Tau exhibits compelling hints of substructures analogous to those being found in other disks \citep{Bouvier_1999}. AA Tau is considered the archetype of a class of young stars with a peculiar form of inner disk driven photometric variability \citep[e.g.][]{McGinnis_2015, Sousa_2016}. While investigating the effects of magnetic fields on inner disk accretion flows, \cite{Bouvier_1999} discovered that AA Tau has photometric variations with an 8.5 day period, similar to the stellar rotation period. As the AA Tau system was thought to be viewed at a high inclination \citep{Basri_1989, Shevchenko_1991, Kwan_1997}, Bouvier et al. suggested that the light curve could be explained by periodic occultation of the star by a warped inner disk. This odd light curve has since motivated intense multi-wavelength scrutiny of AA Tau \citep[e.g.][]{Menard_2003, Andrews_2007, Schmidtt_2007, Oberg_2010, Cox_2013, Zhang_2015b}. In 2011, a substantial dimming ($\sim$2 mag, \textit{V}-band) of the system was also observed, accompanied by significant reddening in the near-IR \citep[$\sim$3-4 mags of visual extinction,][]{Bouvier_2013}. The system has not emerged from this state since \citep{Rodriguez_2015}. 
    
    A possible interpretation of these optical light variations is an extended non-axisymmetric feature (such as a disk warp or protoplanet) passing in front of the star at a distance of \textgreater8 AU, assuming a distance of 145~pc to AA Tau \citep{Bouvier_2013, Rodriguez_2015}. This naturally suggests an investigation of the interplay between fine-scale structures in the outer disk and the inner disk morphology. High resolution ALMA observations of the mm dust disk are therefore an important first step in directly observing such phenomena.
    
    In this paper, we present ALMA observations of the disk around AA Tau that identify it for the first time as multi-ringed, with its mm dust inclination differing substantially from previously inferred inner disk and scattered light inclinations. \S2 describes the details of the observations and the data reduction. \S3 presents the imaged data and an analysis of the visibilities. In \S4, we investigate the discrepancy between our derived inclination and previous measurements, and speculate on the interplay between the outer disk observations and possible inner disk structures.

\section{Observations}
    AA Tau was observed on 2015 July 25 in Band 7 and on 2015 Sept. 29 in Band 6 as part of the ALMA cycle 2 project 2013.1.01070.S. Band 7 observations included 35 antennas with projected baseline lengths between 12 and 1455~m (11-1410~k$\lambda$). The total on-source integration time was 56 minutes. The correlator setup included a Time Division Mode (TDM) continuum window centered at 278.0~GHz with a bandwidth of 2~GHz, as well as two continuum chunks in Frequency Division Mode (FDM) spectral windows, centered at 279.5~GHz and 288.2~GHz with bandwidths of 234 and 469~MHz, respectively. The total continuum bandwidth was 2.7~GHz.
    
    Band 6 observations included 30 antennas with projected baseline lengths between 42 and 2065~m (36-1850~k$\lambda$). The total on-source integration time was 61 minutes. The correlator setup included two TDM mode continuum windows centered at 253.3 and 269.3~GHz, each with a bandwidth of 2~GHz, and an FDM spectral window centered at 255.5~GHz with a bandwidth of 469~MHz. The total continuum bandwidth was 4.47~GHz. HCO$^+$ 3--2 was targeted in an FDM spectral window centered at 267.6~GHz, with a channel spacing of 61~kHz (0.068~km s$^{-1}$).
    
    Separate spectral line Band 7 observations targeting $^{13}$CO in AA Tau were taken on 2016 July 21 as part of the ALMA cycle 3 project 2015.1.01017.S. These observations included 39 antennas, with projected baseline lengths between 12 and 1028~m (13-1170~k$\lambda$). The total on-source integration time was 45 minutes. An FDM spectral window was centered on $^{13}$CO 3--2 at 330.6~GHz.
    
    For the Band 7 continuum observations, the quasar J0423-0120 was used for both phase and bandpass calibration and the quasar J0510+1800 was used for flux calibration. For the Band 6 observations, J0510+1800 was used for phase and bandpass calibration, and J0423-0120 was used for flux calibration. While analyzing the delivered calibrated data, we discovered that the two flux calibrations are in conflict, suggesting an unphysical spectral index of the AA Tau disk ($\alpha\sim$0). The raw Band 7 measurements of J0510+1800 and J0423-0120 (2.3 and 1.0~Jy, respectively) are in good agreement with ALMA calibrator catalog flux values (2.2 and 0.9~Jy, respectively). In contrast, the ratio of the raw Band 6 measurements of J0510+1800 and J0423-0120 (2.1 and 1.0~Jy, respectively) conflicts with the ratio of the catalog values (1.7 and 1.0~Jy, respectively). Due to the narrow spectral windows used for molecular line observations in the correlator setup, phase calibration of the data required mapping of solutions between spectral windows, and an error may have been introduced in this step.
    
    We correct for this conflict by using J0510+1800 as the flux calibrator for both the Band 6 and Band 7 observations, adjusting the flux scaling of the Band 6 observations to correct the J0510+1800 discrepancy (2.1~Jy derived vs 1.7~Jy catalog). The applied flux correction results in a physical spectral index of $\sim$2, but also introduces a substantial uncertainty in the absolute continuum flux of the Band 6 data. While this uncertainty is not problematic when addressing the radial structure of the dust, it does preclude an analysis of the spectral index and thus the grain size distribution across the AA Tau disk. After fixing the flux calibration, we used the disk continuum emission to perform two rounds of phase-only self-calibration in CASA version 4.3.

\section{Results}
    \subsection{Continuum Observations}
\begin{figure*}[ht!]
        \centering
        \includegraphics[width=0.8\textwidth]{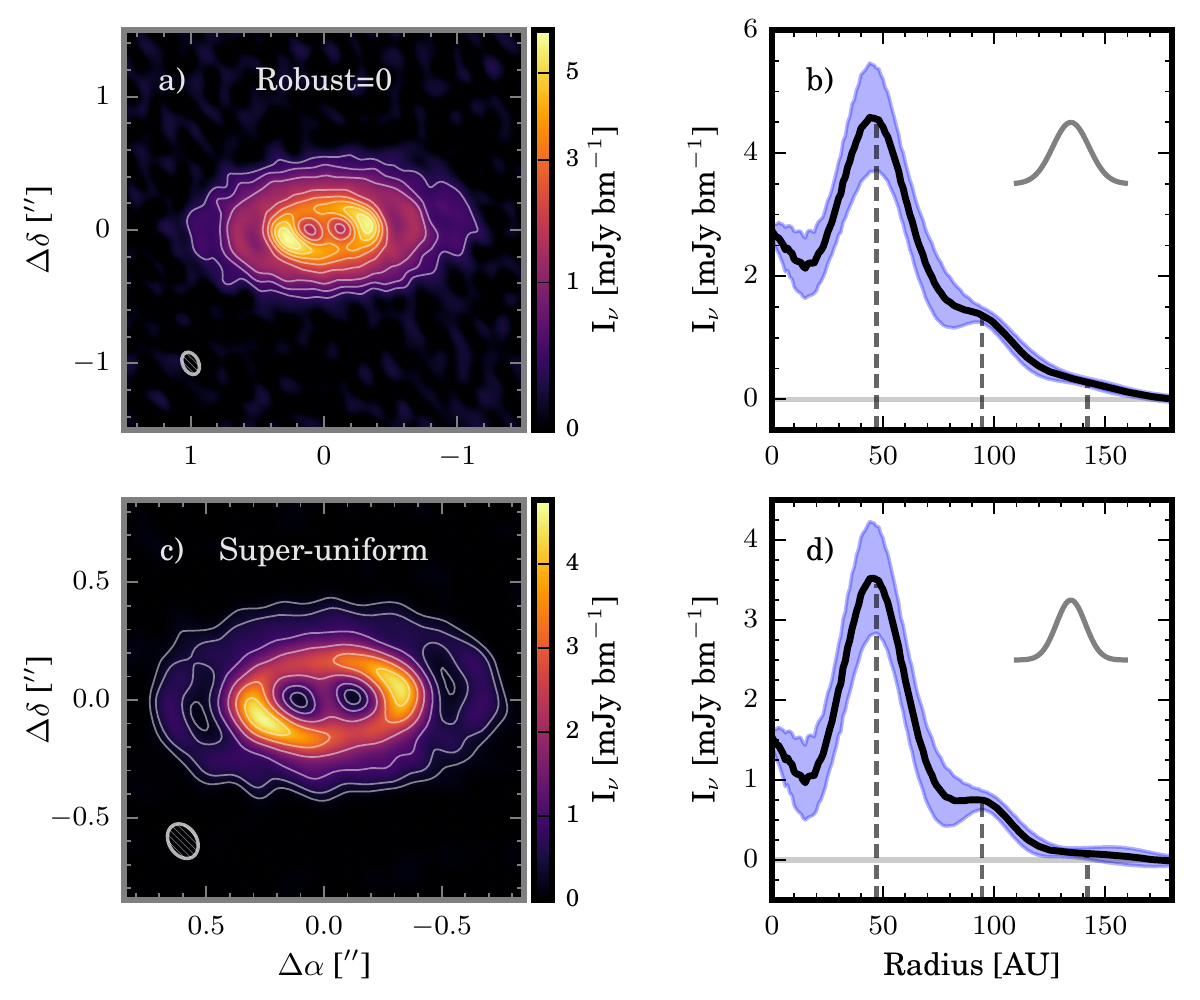}
        \caption{\textit{Panel a:} Synthesized image of the AA Tau dust continuum using combined Band 6 and Band 7 data with a Briggs weighting of robust=0. The beam is 0\farcs18$\times$0\farcs12 and the rms noise is $\sim$44~$\mu$Jy bm$^{-1}$. Contours are $[5,10,20,40,60,...]\times\sigma$. To highlight weaker emission, a power-law stretch has been applied to the color scheme ($\gamma$=0.55). \textit{Panel b:} Deprojected and azimuthally averaged radial profile of the image in panel (a). Data are shown in black and standard deviation of the mean in each radial bin in shaded blue. Black dashed lines denote ring locations derived from our model fit (see \S3.2) and the gray curve represents the synthesized beam. \textit{Panel c:} Same as panel (a), but imaged using a super-uniform weighting to yield a higher spatial resolution. The beam is 0\farcs15$\times$0\farcs11 and the rms noise is $\sim$60~$\mu$Jy bm$^{-1}$. The color scheme is linear. \textit{Panel d:} Same as panel (b), but for the image shown in panel (c). \label{Figure 1}}
        \end{figure*}

        The Band 6 and 7 continuum observations were concatenated and imaged using multi-frequency synthesis \texttt{CLEAN} with a reference frequency of 271.6~GHz. A high sensitivity image of the AA Tau dust continuum (Fig. \ref{Figure 1}, panel a) was created using Briggs `robust' weighting \citep{Briggs_1995}, with a robust parameter of 0, yielding a synthesized beam of 0\farcs18$\times$0\farcs12 (26$\times$17~AU at 145~pc) at a PA of $\sim$27$\degree$ and an rms noise of 44~$\mu$Jy bm$^{-1}$. As seen in the figure, AA Tau hosts a system of nested dust rings with an apparent inclination of $\sim$59$\degree$, which we confirm through visibility modeling (see \S3.2). This is substantially lower than the 71$\pm$1$\degree$ previously fit to infrared scattered light emission \citep{Cox_2013}, and we discuss this discrepancy in \S4.1. Fig. \ref{Figure 1}, panel (b) highlights the location of the three rings in a deprojected and azimuthally averaged radial intensity profile, calculated from the image in panel (a) using the model-constrained inclination of 59$\degree$ and PA of 93$\degree$. The third ring is marginally detected in this radial profile, but confirmed through our visibility modeling (see \S3.2). The three rings are nearly evenly-spaced, peaking at 0\farcs34, 0\farcs66, and 0\farcs99 (49, 95, and 143 AU).
        
        To better display structure in the inner two rings, we additionally imaged the data using super-uniform weighting (Fig. \ref{Figure 1}, panel c), yielding a synthesized beam of 0\farcs15$\times$0\farcs11 (22$\times$16~AU at 145~pc) at a PA of $\sim$33$\degree$ and a slightly higher rms noise of 60~$\mu$Jy bm$^{-1}$. A deprojected and azimuthally averaged radial profile is shown in Fig. \ref{Figure 1}, panel (d). The higher resolution image shows that the inner ring has apparently symmetric azimuthal variations, and there is a `bridge' of emission across the central clearing. At the resolution of the current observations, it is not immediately clear whether this `bridge' is caused by an unresolved inner disk, spiral arms \citep[e.g.][]{Muto_2012, Perez_2016}, or dust streamers similar to those hinted at in HD 142527 \citep{Casassus_2013}. We discuss these possibilities further in \S4.2.
    
    \subsection{Continuum modeling}
        We have interpreted the continuum emission with a simple parametric model composed of three Gaussian rings and a Gaussian inner disk. For each ring, the peak radius, width (FWHM), inclination, position angle, integrated flux in each band, and central offset were allowed to vary. As the observations do not have sufficient spatial resolution to constrain the PA and inclination of the inner disk, we fix the PA to 93$^{\degree}$, constrained by the PA of the observed jet \citep{Cox_2013}, and the inclination to 75$^{\degree}$, constrained by photopolarimetry modeling \citep{OSullivan_2005}. The FWHM and flux of the inner disk were allowed to vary. Two nuisance parameters were added to constrain the central position of the inner disk, which was then treated as the reference point for the outer ring offsets. This 29 parameter model was fit to the observed visibilities using the MCMC routine \texttt{emcee} \citep{Mackey_2013} and the visibility sampling routine \texttt{vis\_sample}\footnote{\texttt{vis\_sample} is publicly available at \url{$https://github.com/AstroChem/vis\_sample$} or in the Anaconda Cloud at \url{$https://anaconda.org/rloomis/vis\_sample$}}, yielding a final best-fit model with a reduced $\chi^{2}$ value of 1.02.

        \begin{table*}
        \begin{center}
        \small
        \begin{threeparttable}[b]
        \caption{Fit model parameters}
        \begin{tabular}{+c^c^c^c^c^c^c^c^c^c^c}
        \toprule
        \otoprule                                                                                
        	                        &   Inner Disk      &   Ring 1          &   Ring 2          &   Ring 3          \\
        	\otoprule
        	r (AU)                  &	--              &   48.7$\pm$0.1    &   94.9$\pm$0.2    &  142.6$\pm$0.6    \\
        	FHWM (AU)               &	5.4$\pm$1.1     &   22.6$\pm$0.2    &   28.7$\pm$0.7    &   26.4$\pm$1.5    \\
        	i ($\degree$)           &	75$^{a}$        &   58.8$\pm$0.1    &   59.0$\pm$0.1    &   59.6$\pm$0.3    \\
        	PA ($\degree$)          &	93$^{a}$        &   94.1$\pm$0.1    &   92.2$\pm$0.1    &   93.3$\pm$0.3    \\
        	Flux Band 6 (mJy)       &	2.1$\pm$0.04    &   45.3$\pm$0.2    &   29.0$\pm$0.4    &   10.2$\pm$0.3    \\
        	Flux Band 7 (mJy)       &	2.2$\pm$0.05    &   61.0$\pm$0.2    &   34.4$\pm$0.4    &    8.5$\pm$0.3    \\
        	$\Delta\alpha$ (AU)     &	--              &    0.0$\pm$0.1    &   -0.3$\pm$0.2    &   -1.4$\pm$0.4    \\  
        	$\Delta\delta$ (AU)     &	--              &   -1.1$\pm$0.2    &   -2.4$\pm$0.2    &   -4.3$\pm$0.4    \\
        \bottomrule
        
        \end{tabular}
        \begin{tablenotes}
        \item[a] Fixed during fit
        \end{tablenotes}
        \end{threeparttable}
        \end{center}
        \end{table*}
        
        The best-fit values and 68\% confidence intervals (1$\sigma$) for all model parameters are presented in Table 1. The retrieved ring radii are 48.7$\pm$0.1, 94.9$\pm$0.2, and 142.6$\pm$0.6~AU, nearly evenly spaced. The FWHM widths of the rings range from 22 to 29~AU, not much larger than the beam along the major axis of the disk, suggesting the individual rings are, at best, marginally resolved, and they may contain further sub-structure. In addition to ring locations and widths, we constrain the ring inclinations to an average of 59.1$\degree\pm$0.3$\degree$, significantly lower than the 71$\degree\pm$1$\degree$ that \cite{Cox_2013} fit to their scattered light observations. In contrast, the fit PAs of the rings are 92-94$\degree$, in good agreement with \cite{Cox_2013} and the predictions of \cite{Menard_2003} from AA Tau's polarization curve. We also find small (\textless~5~AU) mutual offsets between the ring centers, which may be indicative   
        
        \begin{figure*}[ht!]
        \centering
        \includegraphics[width=0.8\textwidth]{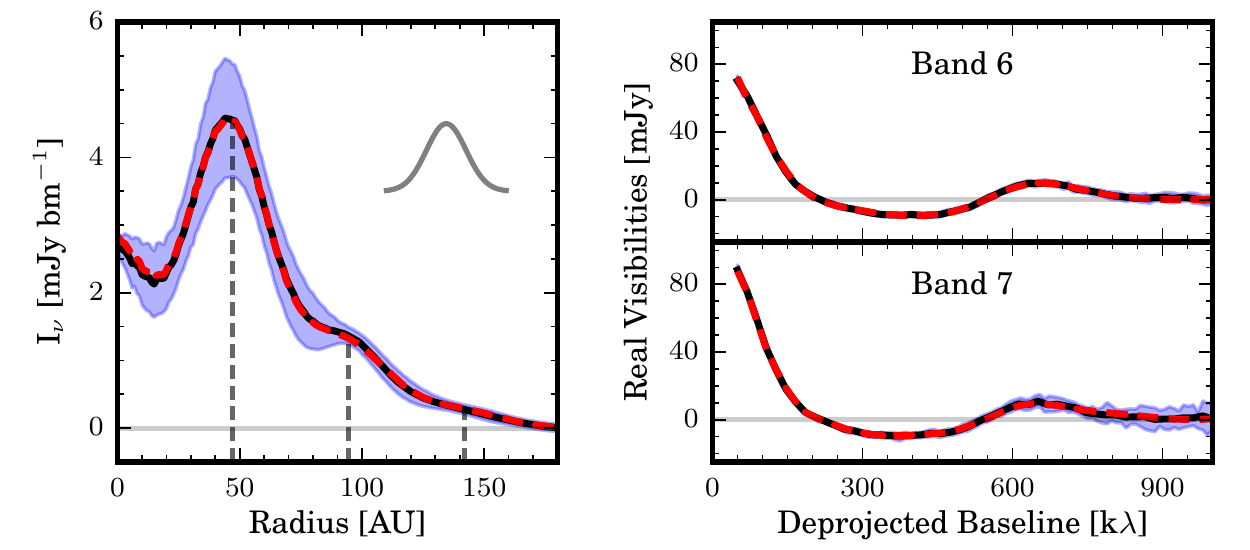}
        \caption{Comparison of dust continuum observations and best-fit model. \textit{Left:} Deprojected and azimuthally averaged radial profile of the imaged continuum in Fig. \ref{Figure 1}, panel (a). Data are shown in black and standard deviation of the mean in each radial bin in shaded blue. Black dashed lines denote ring locations derived from the model fit and the gray curve represents the synthesized beam. Our best fit model is overlaid in dashed red. \textit{Right:} Deprojected real visibilities from the Band 6 and Band 7 continuum observations, binned at 10k$\lambda$ intervals. Data are shown in black, standard deviation of the mean in each radial bin in shaded blue, and the best fit model in dashed red. \label{Figure 2}}
        \end{figure*}   
        
        \noindent{of geometries not considered by our simple model (e.g. eccentricity in the rings).}
        
\begin{figure*}[ht!]
        \centering
        \includegraphics[width=\textwidth]{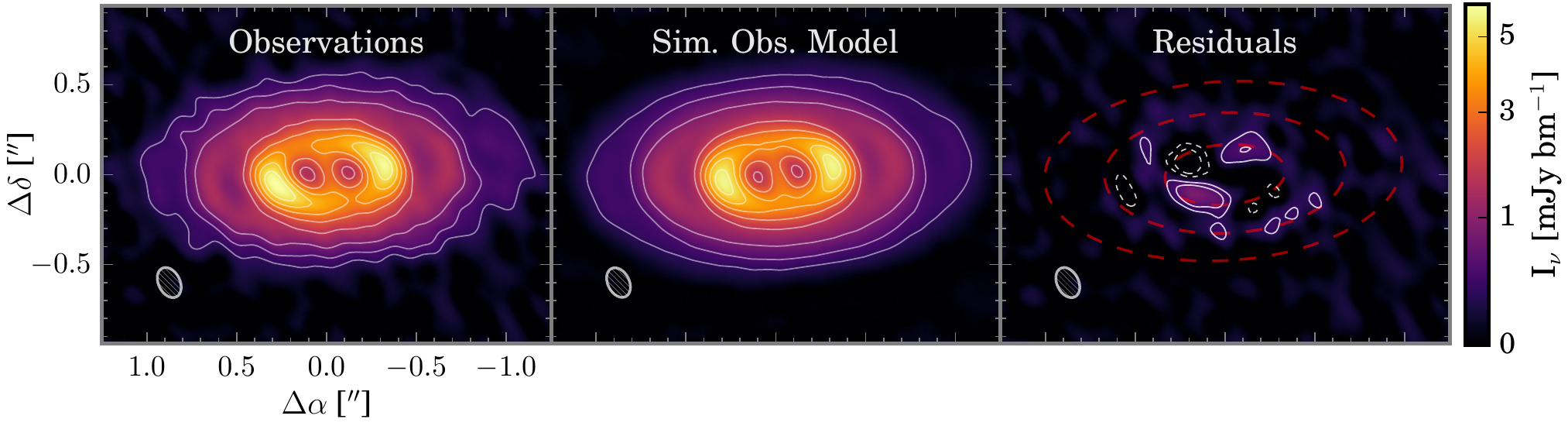}
        \caption{\textit{Left:} Synthesized image of the AA Tau dust continuum using combined Band 6 and Band 7 data with a Briggs weighting of robust=0. The beam is 0\farcs18$\times$0\farcs12 and the rms noise is $\sim$44~$\mu$Jy bm$^{-1}$. Contours for all panels are $[5,10,20,40,60,...]\times\sigma$. \textit{Middle:} Sim-observed model. \textit{Right:} Imaged residuals with ring locations overlaid in dashed red. \label{Figure 3}}
        \end{figure*}

        This parametric model replicates well the observed visibilities and azimuthally averaged radial intensity profile (Fig. \ref{Figure 2}). When comparing the imaged data and simulated observations (using the CASA task \texttt{simobserve}) of our best-fit model, however, it becomes clear that an axisymmetric model does not fit the data perfectly (Fig. \ref{Figure 3}). After subtracting the model visibilities from the data and imaging the residuals, we find structured residuals with a peak of $\sim$12~$\sigma$ ($\sigma$ = 44~$\mu$Jy bm$^{-1}$), suggesting that there is azimuthal structure which is not captured by our models. This is consistent with the symmetric azimuthal variations seen in the imaged data (Fig. \ref{Figure 3}, left). Beam convolution often creates artificial bright spots on either side of an inclined ring, but the simulated observations show that the opposing `tails' of emission cannot be explained by this effect.
        
        To address the origin of these residuals, we tested several variations of our simple concentric ring model. First, we fit a model with independent ring parameters for each of the Band 6 and Band 7 datasets. We found small differences between the bands for all parameters, but they remained broadly consistent with the values in Table 1, and residuals were not improved over the previously described model. As both models produced structured residuals, it is unclear whether the differences between the Band 6 and Band 7 parameters are real. Given this structure in the residuals, we also investigated if a combination of two point sources embedded in the disk near the innermost ring could replicate the observations. We added two variable strength point sources fixed at the innermost ring radius with a variable azimuthal location to the model, and fit it identically to the first model. Residuals improved ($\sim$4$\sigma$), but remained structured, suggesting that the responsible feature is resolved and not point-like.
        
        Finally, we tested if eccentricity could be responsible by adding two model parameters to the innermost ring: an eccentricity and an angle of perihelion. An ellipse with a Gaussian cross-section was used to describe the ring, with a 1/r$^2$ term added to approximately account for pericenter glow \citep[e.g.][]{Wyatt_1999}. After fitting the model using the procedure previously described, we found that a slight amount of eccentricity was preferred for the inner ring, but the residuals barely improved ($\sim$1$\sigma$) and remained structured. We therefore conclude that eccentricity alone cannot explain the observed azimuthal structure. Several other possible explanations are discussed in \S4.2.

    \subsection{Spectral line emission}
        \begin{figure*}[ht!]
        \centering
        \includegraphics[width=0.8\textwidth]{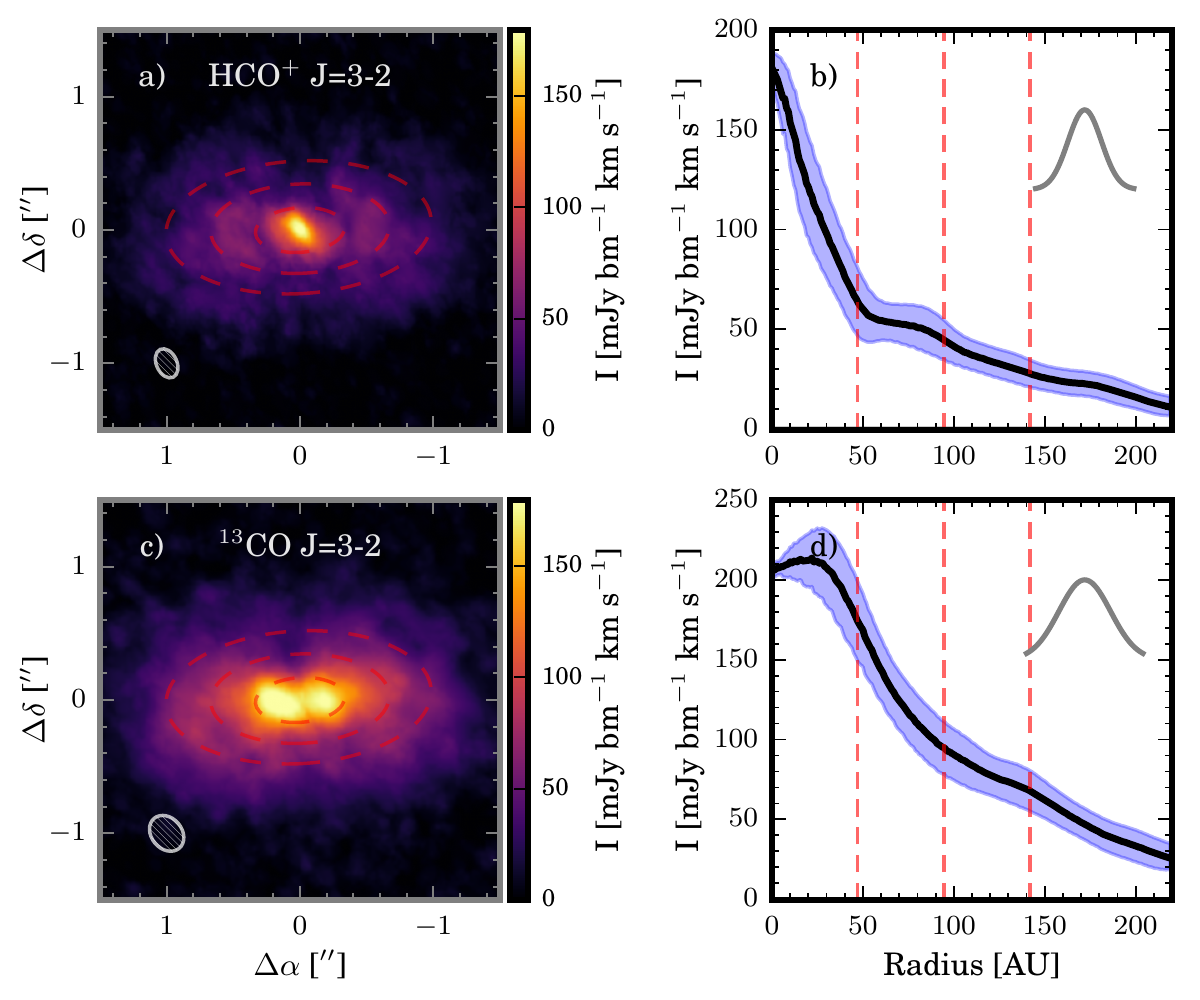}
        \caption{\textit{Panel a:} Moment-0 map showing total integrated HCO$^+$ 3--2 emission. The beam is 0\farcs23$\times$0\farcs15 at a PA of $\sim$25$\degree$ and the rms noise is $\sim$5~mJy bm$^{-1}$ km s$^{-1}$. Locations of the three rings are overlaid in dashed red. \textit{Panel b:} Deprojected and azimuthally averaged radial profile of the HCO$^+$ emission shown in panel (a). Data are shown in black and standard deviation of the mean in each radial bin in shaded blue. Red dashed lines denote ring locations and the gray curve represents the synthesized beam. \textit{Panel c:} Moment-0 map showing total integrated $^{13}$CO 3--2 emission. The beam is 0\farcs29$\times$0\farcs23 at a PA of $\sim$40$\degree$ and the rms noise is $\sim$6~mJy bm$^{-1}$ km s$^{-1}$. Locations of the three rings are overlaid in dashed red. \textit{Panel d:} Same as panel (b), but for the $^{13}$CO emission shown in panel (c). \label{Figure 4}}
        \end{figure*}

        To trace molecular gas kinematics in the disk, we analyzed HCO$^+$ 3--2 emission. The observations were continuum subtracted using \texttt{uvcontsub} and then imaged with \texttt{CLEAN} using natural weighting at a channel resolution of 0.4~km s$^{-1}$. A custom \texttt{CLEAN} mask created for each channel to match the emission, and \texttt{CLEAN}ing was terminated when residuals reached 3$\sigma$. An integrated emission map (Fig. \ref{Figure 4}, panel a) shows the HCO$^+$ emission to be quite bright interior to the innermost ring, compared to the outer disk emission. This is especially clear in a deprojected and azimuthally averaged HCO$^+$ radial profile (Fig. \ref{Figure 4}, panel b), where the emission intensity increases by over a factor of three interior to the innermost ring. In contrast, $^{13}$CO 3--2 emission (Fig. \ref{Figure 4}, panels c~\&~d, identically imaged to the HCO$^+$), is not centrally peaked, and its radial profile is essentially flat within the inner dust ring. The position angle and inclination of the dust continuum were used for deprojection. The radial profiles show kinks near the locations of the inner two rings, but it is unclear whether this reflects gas density variations, dust opacity effects, or artifacts of deprojecting flared emission with a single inclination.
        
\begin{figure}[ht!]
        \centering
        \includegraphics[width=0.47\textwidth]{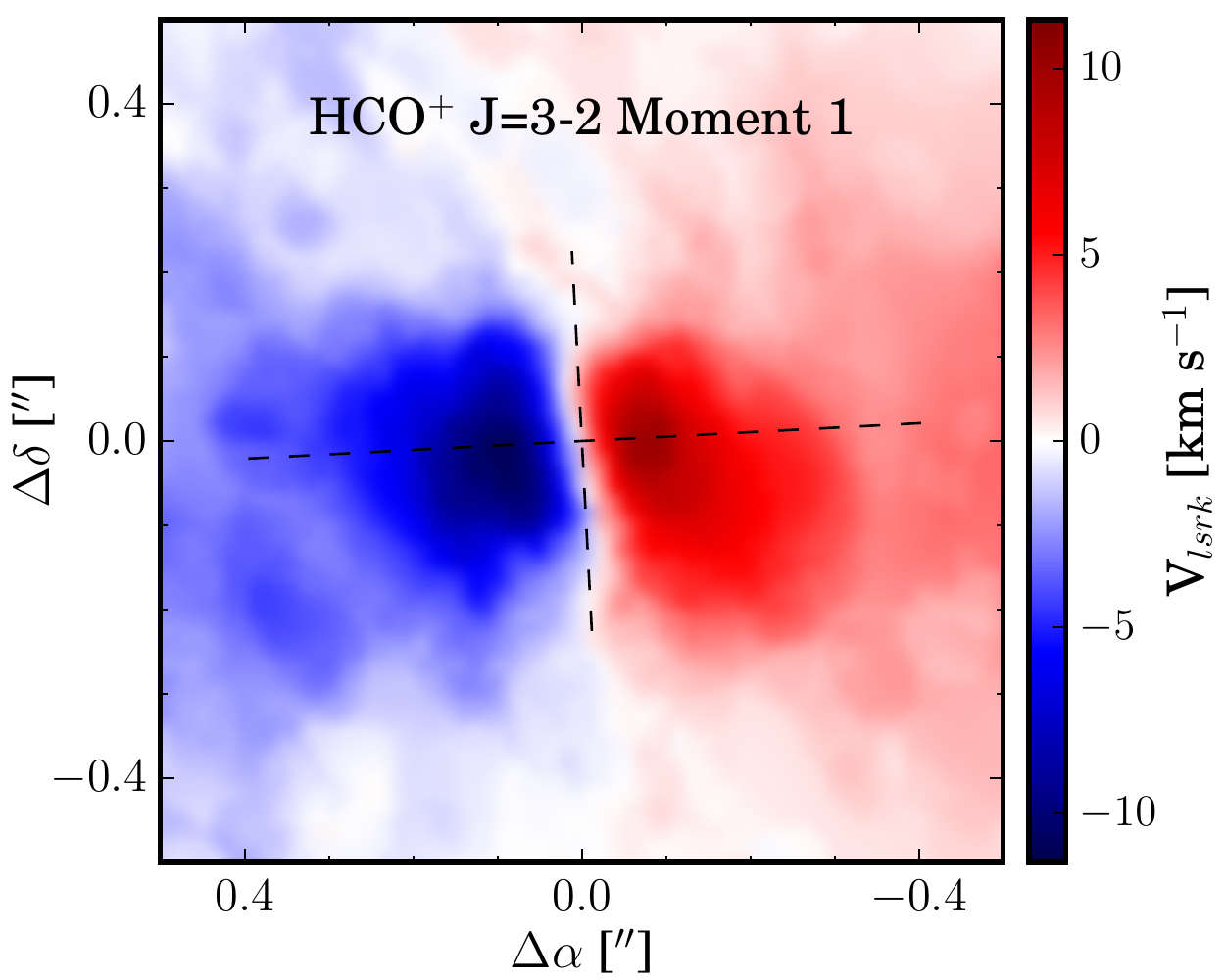}
        \caption{Moment-1 map of HCO$^+$. The dust continuum orientation is shown in dashed black.\label{Figure 5}}
        \end{figure}

        In any case, the presence of sharply centrally peaked HCO$^+$ emission and flat $^{13}$CO emission interior to the inner dust ring is interesting, as it implies that a high bulk gas density (traced by $^{13}$CO) interior to the ring is not responsible for the HCO$^+$ emission intensity. The bright inner disk emission must therefore result from peculiar excitation effects or enhanced HCO$^+$ formation chemistry (e.g. from a high-ionization environment). The HCO$^+$ moment-1 map (Fig. \ref{Figure 5}) is similarly intriguing, showing a twist interior to the inner dust ring, with the emission misaligned by $\sim$10$\degree$ with respect to the continuum orientation. This feature is similar to previous observations of CO in HD 142527 and HD 100546 \citep{Rosenfeld_2014, Pineda_2014} and HCO$^+$ in HD 97048 \citep{vanderPlas_2016, Walsh_2016}, and has been suggested to indicate either a disk warp or radial flow. 
        
        \begin{figure}[ht!]
        \centering
        \includegraphics[width=0.47\textwidth]{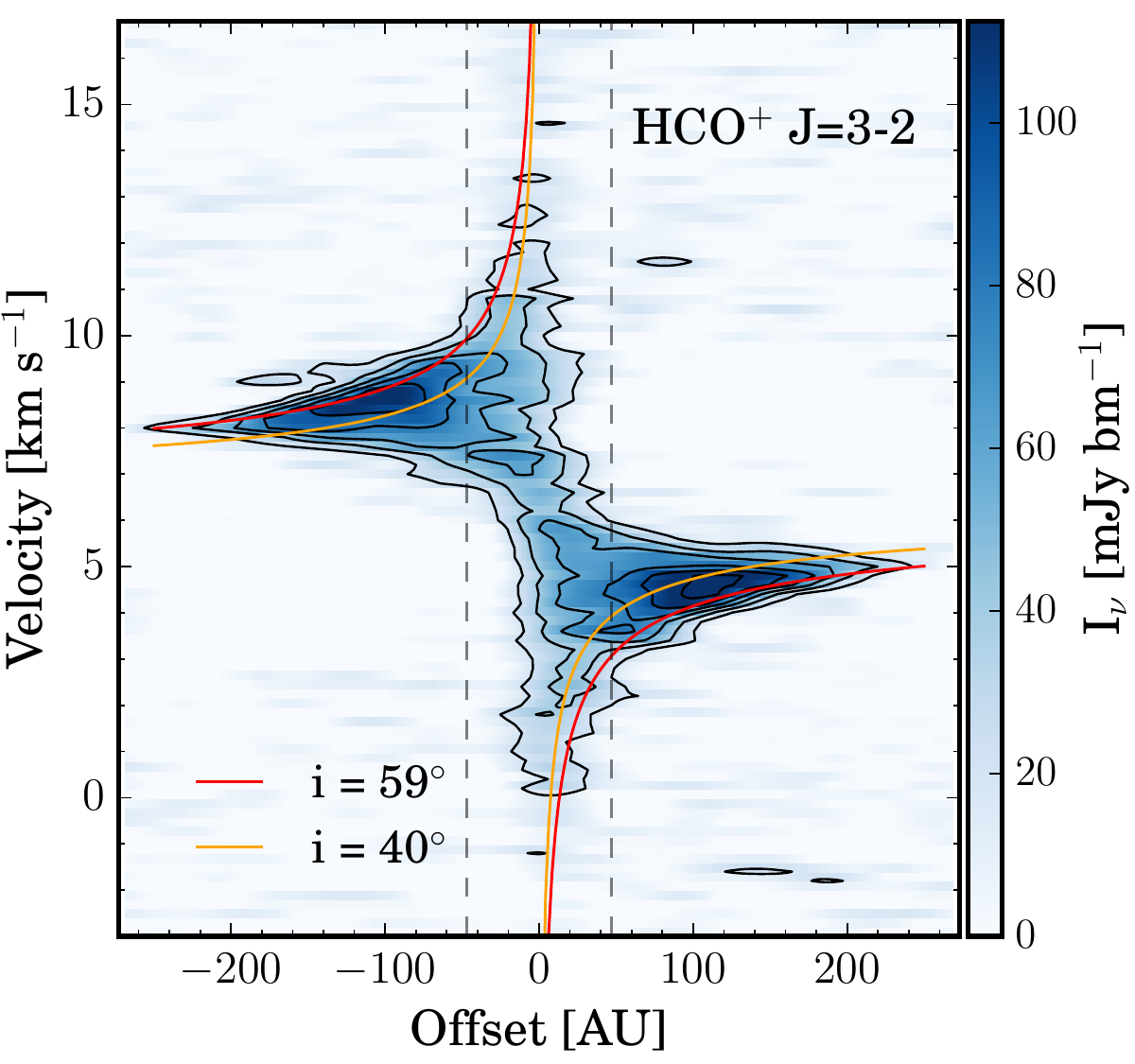}
        \caption{Position-velocity diagram of HCO$^+$ 3--2 emission around AA Tau, modeled after \cite{Pineda_2014}. Contours are $[3,6,9,...]\times\sigma$, $\sigma$~=~6.5~mJy bm$^{-1}$. The expected Keplerian velocity profiles for two disk inclinations (59$\degree$ and 40$\degree$) and a stellar mass of 0.85 M$_{\odot}$ are shown in red and orange, respectively. The location of the first dust ring is denoted by the gray dashed line. \label{Figure 6}}
        \end{figure}
        
        A position-velocity diagram of the HCO$^+$ emission (Fig. \ref{Figure 6}) provides additional evidence for kinematics which cannot arise from a single Keplerian velocity field. The expected Keplerian velocity profiles for two inclinations (59$\degree$ and 40$\degree$) and a stellar mass of 0.85 M$_{\odot}$ are overlaid in red and orange, respectively. Although the higher inclination of 59$\degree$ is able to describe the HCO$^+$ emission well in the outer regions of the disk, it is not consistent with the emission inside of the innermost ring, which is better fit by a lower inclination of 40$\degree$. Similar to the signature in the moment-1 map, this could be indicative of either a disk warp or a non-Keplerian radial flow \citep{Pineda_2014}. We discuss possible interpretations of these kinematic signatures in \S4.2.

\section{Discussion}
    \subsection{Dust rings and inclination}
        We have found that AA Tau hosts three nearly evenly-spaced mm dust rings at an inclination of $\sim$59$^{\circ}$, adding it to a growing list of substructured disks with rings and gaps. In contrast to HL Tau and TW Hya \citep{ALMA_2015, Andrews_2016}, however, which host power law dust disks with numerous narrow gaps, the dust in AA Tau is distributed in rings with broad gaps (Fig. \ref{Figure 1}, panel c), more similar to HD 163296 and HD 169142 \citep{Isella_2016, Fedele_2017}. If the dust gaps in AA Tau result from a planet-disk interaction, as suggested for HD 163296 and HD 169142, multiple massive planets might be involved \citep{Pinilla_2012, Picogna_2015}, although \cite{Gonzalez_2015} have shown that a single massive planet can also create multiple outer dust rings. More detailed modeling is necessary to interpret the observations in this vein, however, as our simple parametric model describes only the continuum surface brightness, rather than the disk surface density.
        
        The inclination (59.1$\pm$0.3$\degree$) of the outer disk dust rings we observe significantly deviates from the inclinations of both the scattered light disk \cite[71$\pm$1$\degree$,][]{Cox_2013} and the inner disk \cite[$\sim$75$\degree$, ][]{OSullivan_2005}. Understanding the true disk inclination is imperative, as a close to edge-on viewing geometry underpins the warped inner disk explanation of AA Tau's short term variability. This discrepancy suggests that either the previous inclinations were skewed, had larger uncertainties than reported, or the inner and outer disks are misaligned \citep[e.g.][]{Marino_2015}.
        
        \begin{figure*}[ht!]
        \centering
        \includegraphics[width=0.8\textwidth]{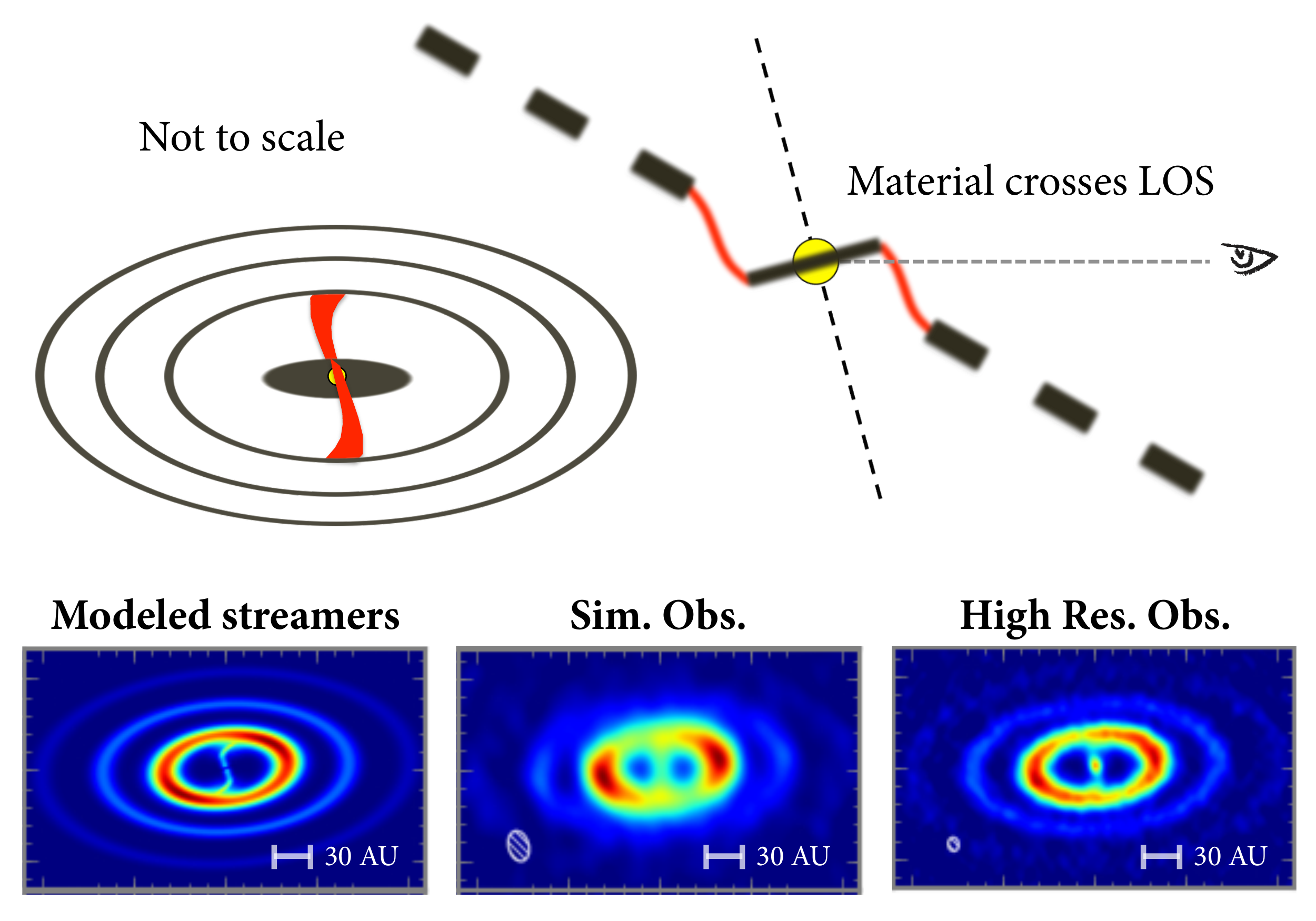}
        \caption{Schematic of possible AA Tau disk and streamer geometry. \textit{Bottom left:} Model modified to include streamers. \textit{Bottom middle:} Observations of streamer model simulated with current resolution. \textit{Bottom right:} Simulated high resolution (0\farcs05) ALMA observations of streamer model. \label{Figure 7}}
        \end{figure*}
        
        Supporting the third possibility, we find an opposite absolute disk orientation compared to \cite{Cox_2013} (i.e. they report an inclined disk with a northern near side, while we find a southern near side), with both orientations being fairly secure. Their orientation is derived both from the scattered light and from the observed jet, which is presumably aligned with the stellar axis and inner disk. In contrast, our orientation is derived from the HCO$^+$ and CO emission geometry in their channel maps (not shown). In moderately inclined disks, \cite{Rosenfeld_2013} have shown that emission arising from a vertically flared $\tau$=1 surface directly traces the absolute disk orientation. Misaligned inner and outer disks therefore appear to be the simplest explanation of all datasets, illustrated in a schematic diagram in Fig. \ref{Figure 7}. Such an orientation would remain consistent with the warped inner disk explanation of AA Tau's periodic photometric variability \citep{Bouvier_1999}, and the deviation between the mm and scattered light inclinations could be explained by shadowing from the inner disk \citep{Dong_2015}.

    \subsection{Non-axisymmetric disk substructure}
        The continuum observations show azimuthal variations in the innermost ring which are resolved and not explained by eccentricity alone. Several non-mutually exclusive possibilities could explain these observations. First, shadowing from a misaligned inner disk could affect the dust temperature, and therefore emission of the inner ring. Second, spiral arms may be present in the disk \citep[e.g.][]{Muto_2012, Perez_2016}. Third, gap crossing streamers could be present in both dust and gas around AA Tau. Dust streamers have been previously suggested in two disks \citep{Casassus_2013, Dutrey_2014}, although the former was not substantiated in further observations \citep{Fukagawa_2013, Muto_2015}. Beam convolution could cause these streamers to manifest as a non-axisymmetric contribution to the inner ring. Due to the suggestive HCO$^+$ gas kinematics observed, we consider here the observable effects of this third scenario.
        
        In general, the need for gap-crossing flows is observationally motivated, as accretion onto the central star continues to be observed even when gaps are present in disks. The small circumstellar disk will be rapidly depleted by accretion unless it is replenished \citep[e.g.][]{Verhoeff_2011}, implying that material must be crossing the inner gap. Models suggest that planets can drive dynamical instabilities which allow material to funnel into gap-bridging filaments \citep[e.g.][]{DodsonRobinson_2011}. Observational evidence for these radial flows is mostly indirect \citep{Beck_2012, Rosenfeld_2014, Zhang_2015b, vanderPlas_2016, Walsh_2016}, but ALMA has begun to allow direct imaging \citep{Casassus_2013, Dutrey_2014}.
        
        Radial gas flows in the AA Tau disk have previously been suggested as an interpretation of infrared CO absorption measurements \citep{Zhang_2015b}. Our HCO$^+$ observations may provide further indirect evidence for such a flow. The `kink' in the HCO$^+$ moment-1 map and the shape of the P-V diagram have both been previously invoked as signatures of radial gas flows and disk warps \citep{Rosenfeld_2014, Pineda_2014}. Furthermore, the extreme brightness and broad line-width of emission within the innermost ring suggests possibly enhanced HCO$^+$ formation in an ionizing environment (e.g. an accretion shock).
        
        If AA Tau does host a radial flow, the central `bridge' and azimuthal asymmetry in the inner ring might then be explained by the presence of dust streamers. The addition of dust streamers to our best fit model, shown in the bottom panels of Fig. \ref{Figure 7}, is consistent with the observed continuum emission within the inner cavity. Such streamers, which shear off from the walls of the inner ring and spiral into the inner disk, are additionally able to replicate the twisted azimuthal variations of the inner ring. Future higher resolution ($\sim$0\farcs05) ALMA Cycle 4 observations (Fig. \ref{Figure 7}, bottom right) will be able to distinguish between this scenario and a cavity with only an inner disk and no streamers.

    \subsection{Relationship between disk structure and photometric variability}
        As previously noted, AA Tau is the archetypal source for a class of stars with similar photometric variability. Recent observations have shown that a number of such `dipper stars' \citep{Ansdell_2016a} host millimeter dust disks at a wide range of inclinations \citep{Ansdell_2016b}, at odds with the warped edge-on inner disk explanation of their variability. As suggested in \cite{Ansdell_2016b}, misaligned inner and outer disks might explain this apparent dilemma, and our observations provide the first evidence for such a geometry in one of these systems. Furthermore, a misaligned inner and outer disk in AA Tau would present the interesting possibility of gap crossing material periodically intersecting our line of sight (Fig. \ref{Figure 4}). This provides a physical motivation for the non-axisymmetric over-density suggested by \cite{Bouvier_2013} and \cite{Rodriguez_2015} to explain the long-duration dimming of AA Tau. If an inner disk radius of several AU is assumed (consistent with our observations), then a direct path for a streamer between the inner ring and inner disk would suggest a LOS crossing distance between 5 and 10~AU, consistent with previous estimates of \textgreater8 AU \citep{Rodriguez_2015}. Higher resolution observations of AA Tau and similar objects will be needed to test both of these hypotheses.
    
\acknowledgments
We would like to thank Ilse Cleeves, Joey Rodriguez, and Andrew Vanderburg for productive discussions. We also thank the anonymous referee for providing comments that greatly improved the quality of the manuscript. RAL and MAM gratefully acknowledge funding from National Science Foundation Graduate Research Fellowships and ALMA Student Observing Support. KIO acknowledges funding from the David and Lucile Packard Foundation. The National Radio Astronomy Observatory is a facility of the National Science Foundation operated under cooperative agreement by Associated Universities, Inc.  This paper makes use of the following ALMA data: ADS/JAO.ALMA\#2013.1.01070.S. ALMA is a partnership of ESO (representing its member states), NSF (USA) and NINS (Japan), together with NRC (Canada) and NSC and ASIAA (Taiwan), in cooperation with the Republic of Chile. The Joint ALMA Observatory is operated by ESO, AUI/NRAO and NAOJ.

\bibliography{refs}

\end{document}